\documentclass[9pt,conference]{IEEEtran}
\usepackage{multirow}
\usepackage[preprint]{waspaa25}

\usepackage{subcaption}
\usepackage{amssymb}
\usepackage{supertabular}
\usepackage{bm}
\usepackage{svg}
\usepackage{amsmath}
\usepackage{acronym}
\usepackage{hyperref}
\usepackage{cleveref}

\crefname{equation}{Eq.}{Eq.}

\title{ReverbMiipher: Generative Speech Restoration meets \\ Reverberation Characteristics Controllability}
\name{Wataru Nakata$^{1,2}\sthanks{Work done during internship at Google}$,
      Yuma Koizumi$^1$,
      Shigeki Karita$^1$,
      Robin Scheibler$^1$,
      Haruko Ishikawa$^1$,\\
      \em Adriana Guevara-Rukoz$^3$,
      Heiga Zen$^1$,
      Michiel Bacchiani$^1$
}
\address{$^{1}$ Google DeepMind, Tokyo, Japan \;
$^{2}$The University of Tokyo, Tokyo, Japan\;
$^{3}$Google, Zurich, Switzerland\\
wnakata@google.com, koizumiyuma@google.com
}

\newacro{SP}[SP]{spatial perception}
\newacro{AR}[AR]{augmented reality}
\newacro{VR}[VR]{virtual reality}
\newacro{TTS}[TTS]{Text-To-Speech}
\newacro{SSL}[SSL]{self-supervised learning}
\newacro{RIR}[RIR]{room impulse response}
\newacro{SR}[SR]{speech restoration}
\newacro{SE}[SE]{speech enhancement}
\newacro{RT}[RT]{reverberation time}
\newacro{MCD}[MCD]{mel-cepstrum distortion}
\newacro{GPE}[GPE]{gross-pitch error}
\newacro{SPK-sim}[SPK-sim]{speaker similarity}
\newacro{DRR}[DRR]{direct-to-reverberant ratio}

\begin{document}
\maketitle

\begin{abstract}
Reverberation encodes spatial information regarding the acoustic source environment, yet traditional Speech Restoration (SR) usually completely removes reverberation. We propose \textit{ReverbMiipher}, an SR model extending parametric resynthesis framework, designed to denoise speech while preserving and enabling control over reverberation. ReverbMiipher incorporates a dedicated ReverbEncoder to extract a reverb feature vector from noisy input. This feature conditions a vocoder to reconstruct the speech signal, removing noise while retaining the original reverberation characteristics. A stochastic zero-vector replacement strategy during training ensures the feature specifically encodes reverberation, disentangling it from other speech attributes. This learned representation facilitates reverberation control via techniques such as interpolation between features, replacement with features from other utterances, or sampling from a latent space. Objective and subjective evaluations confirm ReverbMiipher effectively preserves reverberation, removes other artifacts, and outperforms the conventional two-stage SR and convolving simulated room impulse response approach. We further demonstrate its ability to generate novel reverberation effects through feature manipulation.
\end{abstract}
\section{Introduction}

Reverberation, a phenomenon originating from the complex interplay of sound reflections and diffractions within an enclosed space~\cite{room_acoutics}, encodes spatial information regarding the acoustic source environment that is fundamental to spatial perception~\cite{bailey2017effect,begault2001direct}. Consequently, the accurate rendering of reverberation is critical for enhancing the immersive quality of technologies such as Augmented Reality and Virtual Reality~\cite{bailey2017effect,larsson2010auditory}.

As generative models continue to demonstrate expanding capabilities in creating diverse media content~\cite{rombach2022high,borsos2023audiolm}, the need for speech processing methods that can preserve and control the spatial properties encoded in reverberation is becoming increasingly important. Historically, however, speech denoising research has primarily focused on removing reverberation, often treating it as undesirable noise that negatively impacts speech intelligibility~\cite{dereverb_book,dereverb1,reverb_challenge,dereverb2,dereverb3,das2021fundamentals}. Recent advancements show that modern \ac{SR} techniques are capable of transforming noisy, reverberant recordings into high-fidelity, studio-quality audio~\cite{Maiti_waspaa_2019,self_remaster,Su_2021,liu22y_interspeech,UNIVERSE,koizumi_2023_sr,Richter_2023,diffusion_sr_review_2024,scheibler24_interspeech,kang2025,genhancer,liu2024jointsemanticknowledgedistillation,ditse,miipher2}. Considering the success of generative models in various speech generation tasks~\cite{borsos2023audiolm}, it is logical to pursue generative speech restoration models capable not only of cleaning audio but also of actively preserving and controlling reverberation characteristics. Such models would offer creation of realistic and diverse datasets for audio spatial understanding models. Furthermore, they would enable dynamic reverberation preservation or manipulation, offering fine-grained control over the acoustical environment in recent video generation models~\cite{gupta2023,blattmann2023stable}.

Regarding reverberation control, room acoustics simulation has remained a field of active research since the 1960s, spawning a wide array of geometric~\cite{image_method,10.1121/10.0023935}, wave-based~\cite{fdtd_rir}, and deep learning techniques~\cite{luo2022learning,irgan,fins}. One potential approach for our task involves deriving a dry source signal using SR, followed by a convolution with a simulated \ac{RIR}. However, this two-stage process faces limitations, notably the difficulty of explicit RIR estimation, and the susceptibility to error propagation, potentially compromising output audio fidelity. Consequently, end-to-end architectures that directly manage reverberation offer a potentially more robust alternative, drawing parallels with successful fully neural~\cite{choi23e_interspeech,he2025multi} or language model-based~\cite{borsos2023audiolm} systems used in fields like speech synthesis and conversion.

We propose \textit{ReverbMiipher}, an SR model based on Miipher-2\cite{miipher2}, designed to preserve and manipulate reverberation while removing noise. Extending Miipher-2's parametric resynthesis approach, ReverbMiipher incorporates a ReverbEncoder network that extracts a dedicated reverb-feature vector encoding reverberation information from the noisy input. Conditioning a vocoder with this feature yields a denoised waveform that preserves the original reverberation. To ensure this feature captures only reverberation, training involves stochastically replacing it with a zero-vector, directing the model to output the anechoic source waveform under this condition. This architecture enables reverberation control by manipulating the reverb feature, e.g., via replacement using features from other utterances, interpolation, or sampling from a learned latent space.
Objective and subjective evaluations demonstrate that ReverbMiipher removes artifacts while preserving reverberation, outperforming the two-stage approach, i.e. SR output convolved with simulated RIRs. We also demonstrate generation of novel reverberation via interpolation or principal component analysis (PCA)-based latent space sampling of the model's reverb features.
Speech samples are available online\footnote{
    \url{https://google.github.io/df-conformer/reverb_miipher/}
\label{fot:demo}
}. %

\section{Conventional Methods}

\subsection{Generative speech restoration}

In this paper, SR is defined as the task of estimating a high-quality signal by removing all forms of speech degradation, including but not limited to enhancement~\cite{enhancement}, dereverberation~\cite{dereverb_book,dereverb1,reverb_challenge,dereverb2,dereverb3,das2021fundamentals}, declipping~\cite{zavivska2020survey}, and super-resolution~\cite{li2015deep,kuleshov2017audio}. This task is also referred to in other literature as ``Universal Speech Enhancement''~\cite{urgent} or ``High-Fidelity Speech Enhancement''~\cite{genhancer}.
Let the $T$-sample time-domain signal $\bm{x} \in \mathbb{R}^T$ be a degraded observation of an original signal $\bm{s} \in \mathbb{R}^{T}$.
The goal of SR is to find a function $\mathcal{F}$ that estimates $\bm{s}$ from $\bm{x}$ as $\bm{y}= \mathcal{F}(\bm{x}) \in \mathbb{R}^{T}$.

Current trends in SR research indicate a shift from paradigms based on the modification of $\bm{x}$, such as time-frequency masking~\cite{enhancement}, to those focused on the direct synthesis of $\bm{s}$ utilizing generative models~\cite{Maiti_waspaa_2019,self_remaster,Su_2021,liu22y_interspeech,UNIVERSE,koizumi_2023_sr,Richter_2023,diffusion_sr_review_2024,scheibler24_interspeech,kang2025,genhancer,liu2024jointsemanticknowledgedistillation,ditse,miipher2}. Generative model-based approaches estimate the probability distribution of $\bm{s}$ conditioned on $\bm{x}$. The output signal is then obtained by sampling from this estimated distribution, i.e. $\bm{y} \sim p(\bm{s} \mid \bm{x})$. A potential consequence of this waveform generation process is the divergence of speech content or speaker characteristics in $\bm{y}$ from those of $\bm{s}$. To mitigate this, state-of-the-art techniques incorporate conditioning inputs (e.g., on text, speaker ID), employ loss functions to maintain phonemic consistency~\cite{scheibler24_interspeech,liu2024jointsemanticknowledgedistillation}, or leverage \ac{SSL} models for robust feature extraction~\cite{koizumi_2023_sr}.

\begin{figure}[t]
    \centering
    \begin{subfigure}[b]{0.351\linewidth}
        \includegraphics[width=\linewidth]{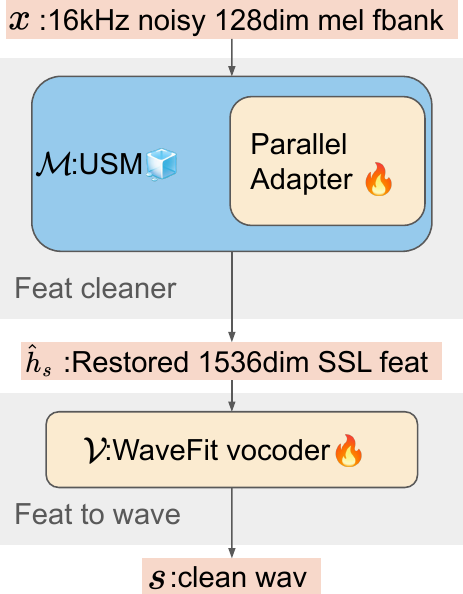}
        \caption{Miipher-2}
        \label{fig:subfig1}
    \end{subfigure}
    \begin{subfigure}[b]{0.531\linewidth}
        \includegraphics[width=\linewidth]{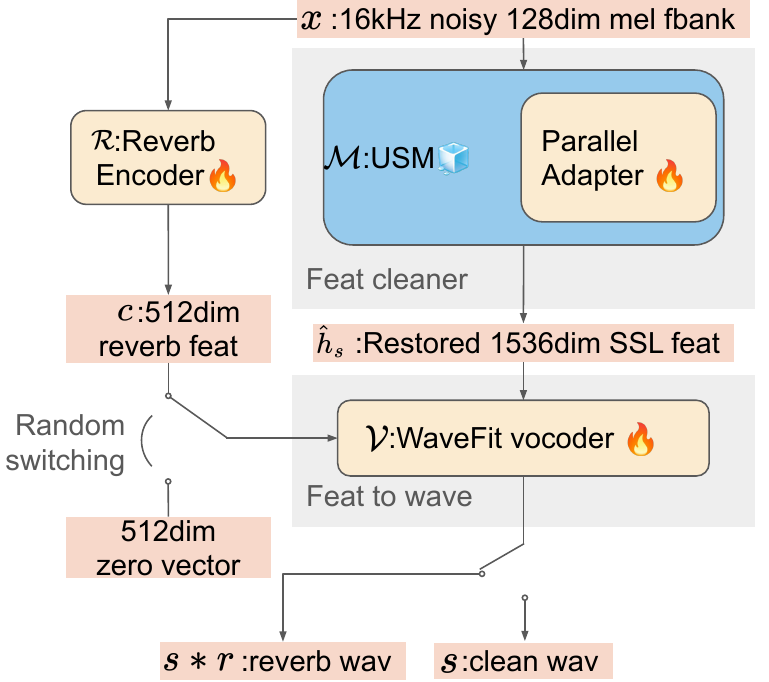}
        \caption{ReverbMiipher}
        \label{fig:subfig2}
    \end{subfigure}
    \caption{Overview of (a) Miipher-2 and (b) the proposed ReverbMiipher.}
    \label{fig:overview}
\vspace{-0.5\baselineskip}
\end{figure}

\subsection{Miipher-2}\label{sec:miipher2}

Miipher-2~\cite{miipher2} is a generative SR model based on a parametric resynthesis strategy~\cite{Maiti_waspaa_2019}. An overview of Miipher-2 is shown in \cref{fig:subfig1}, which comprises two primary components: a feature cleaner, which predicts acoustic features corresponding to a clean waveform from an input noisy waveform, and a vocoder, which subsequently synthesizes a waveform from these predicted clean features.

First, the feature cleaner predicts an \ac{SSL} feature corresponding to clean speech signal $\bm{h}_{s} = \mathcal{M}(\bm{s})$ from a given 128 mel-bands log-mel spectrogram of $\bm{x}$. Miipher-2 uses pre-trained Universal Speech Model (USM)\cite{zhang2023google} as $\mathcal{M}$.
To predict $\bm{h}_{s}$ from $\bm{x}$, Miipher-2 connects parallel adapters~\cite{paralleladapter} to each Conformer~\cite{conformer} layer in USM as $\mathcal{M}_{a}$, i.e $\hat{\bm{h}}_{s} = \mathcal{M}_{a}(\bm{x})$. 
During training only the parallel adapter layers are updated.
The loss function is a sum of mean squared error, mean absolute error and a spectral convergence loss which can be expressed as
$\mathcal{L}_{\mathcal{M}_a} =
\lVert \bm{h}_{s} - \hat{\bm{h}}_{s} \rVert_1
+
\lVert \bm{h}_{s} - \hat{\bm{h}}_{s}\rVert_2^2
+
\lVert \bm{h}_{s} - \hat{\bm{h}}_{s}\rVert_2^2
/
\lVert \bm{h}_{s} \rVert_2^2$.

Then, the neural vocoder converts $\hat{\bm{h}}_{s}$ to $\bm{y} = \mathcal{V}(\hat{\bm{h}}_{s})$. As the neural vocoder, Miipher-2 uses WaveFit~\cite{wavefit} which is inspired by diffusion models and fixed-point iteration. It transforms random numbers into a waveform conditioned on input features, therefore, by conditioning it by $\hat{\bm{h}}_{s}$ this process can be regarded as sampling from $p(\bm{s} \mid \bm{x})$.
First, $\mathcal{V}$ is pre-trained by using ground truth SSL features to minimize
$\mathcal{L}_{\mathcal{V}} (\bm{s}, \mathcal{V}(\bm{h}_{s}))$ where $\mathcal{L}_{\mathcal{V}}$ is the loss function of WaveFit.
Then, $\mathcal{V}$ is finetuned using the predicted SSL feature to minimize
$\mathcal{L}_{\mathcal{V}} (\bm{s}, \mathcal{V}(\hat{\bm{h}}_{s}))$.
For the details of the implementation, please refer to \cite{miipher2}.

\section{ReverbMiipher}

\subsection{Model overview}

The proposed method, ReverbMiipher, is an extended model of Miipher-2 endowed with reverberation controllability as shown in \cref{fig:subfig2}. This model incorporates an additional feature extraction module termed ReverbEncoder $\mathcal{R}$. The ReverbEncoder takes a log-mel spectrogram extracted from $\bm{x}$ and outputs a reverb-feature $\bm{c} = \mathcal{R}(\bm{x}) \in \mathbb{R}^{D}$, which is a vector embedded with reverberation characteristics. Here, $D=512$ is the dimension of the reverb-feature. By using $\bm{c}$ as an additional conditioning for $\mathcal{V}$, ReverbMiipher outputs a signal in which only reverberation is preserved and all other artifacts are removed as $\mathcal{V}(\hat{\bm{h}}_{s}, \bm{c})$. Furthermore, applying control to $\bm{c}$ enables the manipulation of reverberation.

\subsection{ReverbEncoder training}

ReverbEncoder $\mathcal{R}$ consists of a stack of four Conformer~\cite{conformer} layers with a kernel size of five in the convolution module, followed by mean pooling along the time axis. The input of $\mathcal{R}$ is a 128-dimensional log-mel filterbank, which is also used by $\mathcal{M}_{a}$.

A challenge in training the ReverbEncoder is to disentangle reverberation from other acoustic information. If this disentanglement is not achieved, manipulating or interpolating $\bm{c}$ to control the reverberation of the output signal will also alter information such as speaker identity and speech content.
To address this, a stochastic switching technique was introduced during the training of both $\mathcal{R}$ and $\mathcal{V}$. In the creation of supervised training data, the input waveform $\bm{x}$ was generated from the clean speech signal $\bm{s}$ according to
$
\bm{x} = \mathcal{A}\left(
\bm{s} * \bm{r} + \bm{n}
\right),
$
where 
$*$ is convolution,
$\bm{r}$ is an RIR,
$\bm{n} \in \mathbb{R}^T$ is a noise waveform,
and $\mathcal{A}$ signifies a sequence of additional artifact operators, such as codec artifacts and band-width filters. This switching between the reverberant target and the clean speech target is performed independently for each sample within a mini-batch. Consequently, a mini-batch comprises samples targeting both clean and reverberant speech. $\mathcal{R}$ and $\mathcal{V}$ are then trained by minimizing the loss function:
\begin{equation}
    \mathcal{L}_{\mathcal{R}} = \begin{cases}
        \mathcal{L}_{\mathcal{V}} \left(
        \bm{s} * \bm{r}, \mathcal{V}(\hat{\bm{h}}_{s}, \bm{c})
        \right)
        & \text{if $\mathcal{U}(0,1) > q$} \\
        \mathcal{L}_{\mathcal{V}} \left(
        \bm{s}, \mathcal{V}(\hat{\bm{h}}_{s}, \bm{0})
        \right)
        & \text{otherwise } \label{eq:switching}
    \end{cases},
\end{equation}
where
$\mathcal{U}(0,1)$ is a uniform random number with values between 0 and 1,
$\bm{0}$ is a $D$-dimensional zero-vector,
and the hyperparameter $q=0.1$ is the switching probability of the loss function.

The feature cleaner $\mathcal{M}_{a}$ is trained using the Miipher-2's (\cref{sec:miipher2}) methodology. This training approach assumes that the estimated clean speech features  $\hat{\bm{h}}_{s}$ exclusively contain speech information, devoid of reverberation, noise, or other artifacts. By defining the model's output as $\bm{s}$ when the reverberation conditioning vector is $\bm{0}$, the model is explicitly trained to interpret a zero-vector as the absence of reverberation, enabling speech reconstruction solely from  $\hat{\bm{h}}_{s}$. Furthermore, given the target signal is $\bm{s} * \bm{r}$ when the reverberation feature is input to the network, any artifacts present in $\bm{c}$ other than reverberation will be treated as noise by the network. Consequently, the ReverbEncoder weights are expected to be updated to selectively preserve information pertaining only to reverberation.

\section{Experiments}

We demonstrate the efficacy of the proposed method in: (i) preserving reverberation from noisy signals while simultaneously removing other noise and artifacts; and (ii) enabling the manipulation of reverberation through the reverb-feature. The former claim (i) is substantiated in this section through a comparative evaluation of the proposed method with conventional techniques, employing both objective and subjective assessments. The latter claim (ii) is exemplified via case studies in the sections that follow.

\subsection{Comparison method}

To our knowledge, the task of reverberation-controllable SR has not been extensively investigated in prior research. For instance, speech enhancement models primarily designed for denoising may preserve reverberation in the output signal but are often incapable of mitigating other distortions, such as codec artifacts. While generative SR models can be trained with reverberant speech as the target, this approach lacks explicit control over the reverberation characteristics. Therefore, this paper introduces a baseline system consisting of a conventionally trained SR model, the output of which is subsequently convolved with a synthesized RIR. The Miipher-2 model is utilized as the SR component, given its foundational role in the proposed architecture. This baseline system is henceforth designated as \textbf{Miipher-RIR}.

The two-stage processing pipeline of Miipher-RIR requires the estimation of both the anechoic signal and the RIR from the noisy input. However, estimating the RIR from a noisy signal presents an inherently challenging problem. Therefore, in this study, the RIR is simulated using the image method~\cite{image_method}, under the assumption that reverberation time (RT60) and \ac{DRR} are perfectly known. It should be noted that, as practical estimation of these parameters typically incurs errors, the Miipher-RIR performance metrics reported herein likely represent an upper bound compared to those achievable in real-world scenarios.

Initially, RT60 and DRR values were computed from the ground-truth RIR. Subsequently, twenty RIRs were simulated using the image method implemented in pyroomacoustics simulator~\cite{8461310}, employing the ground-truth RT60 and randomly assigning source and microphone positions within the simulated enclosure. From these simulations, the RIR exhibiting the \ac{DRR} value closest to the target was selected for convolution with the output of the Miipher-2 model. The \ac{DRR} was calculated from the RIR according to the definition provided in~\cite{mack2020single}, assuming a direct sound segment length of 8 ms.

\subsection{Experimental conditions}

\paragraph*{\textbf{Training dataset}}
Training was performed using the FLEURS-R speech corpus~\cite{flerus-r}. FLEURS-R is a large-scale multilingual corpus containing approximately 1,300 hours of speech across 102 languages at a sampling rate of 24kHz. Noisy training samples were created by mixing clean speech from FLEURS-R with noise from the internal dataset at random Signal-to-Noise Ratios (SNRs) ranging from 5 dB to 30 dB. Data augmentation techniques included applying reverberation (using the simulated RIRs by the image method) and simulating various codec artifacts including MP3, Vorbis, A-law, AMR-WB, and OPUS codecs following previous work~\cite{koizumi_2023_sr,miipher2}.

\paragraph*{\textbf{Evaluation dataset}}
Evaluation was conducted on a simulated test set created using real recorded RIRs and noise sources distinct from the training data. The clean speech source for the evaluation set was the EARS dataset~\cite{ears}. Ten random utterances were selected per speaker. The number of total utterances was 1070.
The noise sources for the evaluation set were taken from the WHAM! dataset~\cite{Wichern2019WHAM}. Real-world RIRs were sourced from the MIT Acoustical Reverberation Scene Statistics Survey~\cite{mit_rir}. Test samples were generated by first convolving the clean speech with an RIR from the MIT survey, then adding WHAM! noise at random SNR ranging from -5 dB to 20 dB. Finally, a codec simulation was applied, randomly selecting from MP3, Vorbis, A-law, AMR-WB, and OPUS for each sample.

\paragraph*{\textbf{Optimizer}}:
We used an Adam~\cite{adam} with $\beta_1=0.9, \beta_2 = 0.98$ for the vocoder, and $\beta_1=0.9,\beta_2=0.999$ for the feature cleaner. Learning rate warmup was applied for 5000 steps.
The number of training steps were 2 million steps for ReverbEncoder and vocoder, and 250k steps for feature cleaner. The batch size was 512.

\subsection{Subjective evaluation}
A subjective evaluation was conducted to assess the perceptual similarity of reverberation. Given that evaluating subtle reverberation changes can be challenging and may require audio expertise, a standard Mean Opinion Score test, which commonly solicit overall quality judgments, was deemed unsuitable for crowdsourced human reviewers who might lack this specific expertise. Therefore, a pairwise comparison method with ranking was employed. This approach directs listeners' focus, regardless of expertise, specifically onto reverberant characteristics through direct comparison. Human reviewers first listened to the ground truth reverberant speech, then ranked corresponding samples from Miipher-2, Miipher-RIR, and ReverbMiipher based on reverberation similarity to the ground truth. The presentation order of these 3 samples were randomized. The evaluation involved 557 reviewers, each evaluating a maximum of 7 sample sets, yielding 2,878 total ratings. All human reviewers were required to use headphones.
\begin{table}[t] %
\centering %
\caption{Subjective evaluation results. \textbf{Bold} font shows the best results. $p$-values are calculated using pairwise Wilcoxon test~\cite{conover1999practical} with Bonferroni correction~\cite{dunn1961multiple}.} %
\label{tab:subjective_results} %
\begin{tabular}{lr|rr} %
\toprule
\multirow{2}{*}{Model} & \multirow{2}{*}{Mean rank$\downarrow$} & \multicolumn{2}{c}{$p$-value} \\ %
 & & Miipher-RIR&  ReverbMiipher\\
\midrule
Miipher-2 & 2.50 & $< 10^{-8}$ & $< 10^{-8}$ \\
Miipher-RIR & 2.01 &  - & $< 10^{-8}$ \\
ReverbMiipher & \textbf{1.49} &  - & - \\ %
\bottomrule
\end{tabular}
\end{table}
\cref{tab:subjective_results} details the results of the subjective evaluation. The proposed ReverbMiipher yielded reverberation most similar to the ground truth, relative to the other models with statistical significance. Therefore, ReverbMiipher successfully preserves perceptually similar reverberation in SR.
\subsection{Objective evaluation}
Performance was assessed using objective metrics calculated by comparing the model outputs against the ground truth reverberant speech (i.e. $\bm{s} * \bm{r}$). 
The metrics used were:

\noindent\textbf{Mel-Cepstral Distortion (MCD)}~\cite{mcd}: MCD quantifies the spectral distance between the mel-frequency cepstral coefficients (MFCCs).

\noindent\textbf{Gross Pitch Error (GPE)}~\cite{chu2009reducing}: GPE measures the proportion of voiced frames where the estimated fundamental frequency (pitch) deviates by more than 20\% from $\bm{s} * \bm{r}$, primarily assessing prosodic information preservation.

\noindent\textbf{Speaker Similarity (SPK-sim)}: SPK-sim evaluates the preservation of speaker identity. We computed a cosine similarity of the speaker embedding~\cite{NEURIPS2018_6832a7b2,chen2018sample} extracted from the ground truth reverberant speech and the model.

\begin{table}[t] %
\centering %
\caption{Objective evaluation results. \textbf{Bold} font shows the best results among the restored speech.
} %
\label{tab:objective_results} %
\begin{tabular}{l|rrrr} %
\toprule
Model & \textbf{MCD} ($\downarrow$) & \textbf{GPE} ($\downarrow$) & \textbf{SPK-sim} ($\uparrow$) \\ %
\midrule
Noisy input &  8.52 & 0.09 & 0.79 \\
Miipher-2 &  7.28 & 0.16 & 0.63 \\
Miipher-RIR& 6.76 & 0.19 & 0.71\\
ReverbMiipher (ours)& \textbf{5.26} & \textbf{0.16} & \textbf{0.79} \\ %
\bottomrule
\end{tabular}
\vspace{-0.5\baselineskip}
\end{table}

\begin{figure}[t]
    \centering
    \includegraphics[width=0.8\linewidth]{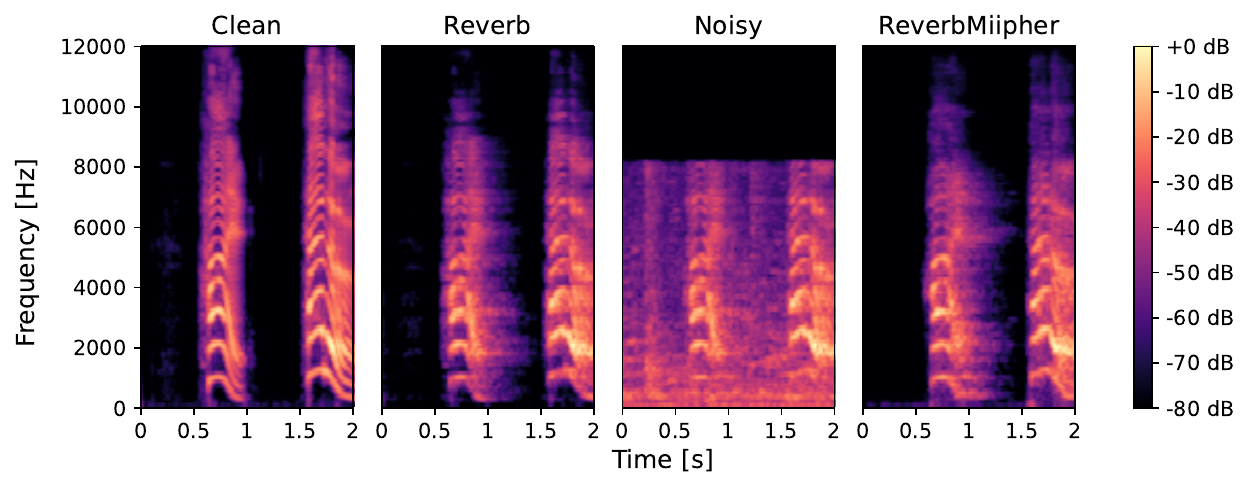}
    \caption{SR result with ReverbMiipher. From the left, the spectrograms of clean signal $\bm{s}$, reverberant target $\bm{s} * \bm{r}$, noisy input $\bm{x}$, and output $\bm{y}$, respectively.}
    \label{fig:sr_result}
\vspace{-0.5\baselineskip}
\end{figure}

The objective evaluation results, including scores for noisy input and Miipher-2~\cite{miipher2} which predicts $\bm{s}$, are presented in \cref{tab:objective_results}. ReverbMiipher achieves the best performance among the three compared models across all three metrics, indicating its success in preserving the reverberance of the input speech. ReverbMiipher outperforms Miipher-RIR.%
Thus, we observe better performance when stochastically integrating the encoding of reverberation within the SR model, as opposed to adding reverberation to clean speech in a post-hoc manner.
The GPE results also suggest that the Miipher-RIR baseline degrades the prosodic information of the original speech. \cref{fig:sr_result} shows an example of restoration result using ReverbMiipher. Although the noisy input contains reverberation as well as noise and the effects of a low-pass filter, we can see that ReverbMiipher removes all signal degradation except for reverberation.

\section{Analysis and Applications}

\subsection{Visualization of reverb-feature distribution}

\begin{figure}
    \centering
    \includegraphics[width=0.85\linewidth]{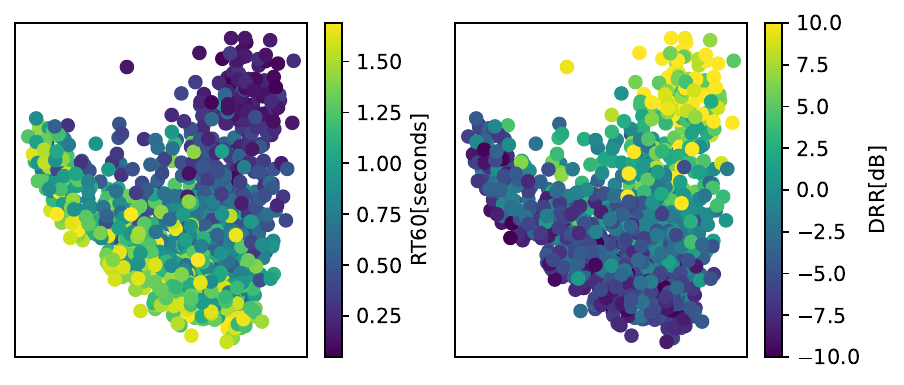}
    \caption{PCA-based latent space distribution of reverb feature with respect to RT60 (left) and DRR (right).}
    \label{fig:reverb_feature}
\vspace{-0.5\baselineskip}
\end{figure}

We investigated the relationship between our proposed reverberation feature, $\bm{c}$, and standard reverberation metrics (RT60 and \ac{DRR}) using PCA visualization. Test signals were created by convolving speech from the EARS corpus with synthetic RIRs generated via pyroomacoustics~\cite{8461310}. For each signal, we extracted $\bm{c}$ and calculated the RT60 (Schroeder method \cite{schroeder1965new}) and \ac{DRR} from RIRs.

\cref{fig:reverb_feature} displays the PCA projection, mapping $\bm{c}$ against RT60 and \ac{DRR}. Results show $\bm{c}$ changes continuously relative to both standard metrics. This smooth variation supports the use of $\bm{c}$ for implementing fine-grained, continuous user control over reverberance during speech restoration. Furthermore, the visualization shows anechoic samples clustering in the upper right, while increasing reverberation causes the data points to spread towards the upper left and lower right. This dispersion highlights the feature's ability to capture the widening range of acoustic properties characteristic of more reverberant environments.

\subsection{Reverb-feature interpolation}

\begin{figure}[t]
    \centering
    \begin{subfigure}[t]{\linewidth}
    \centering
        \includegraphics[width=0.87\linewidth]{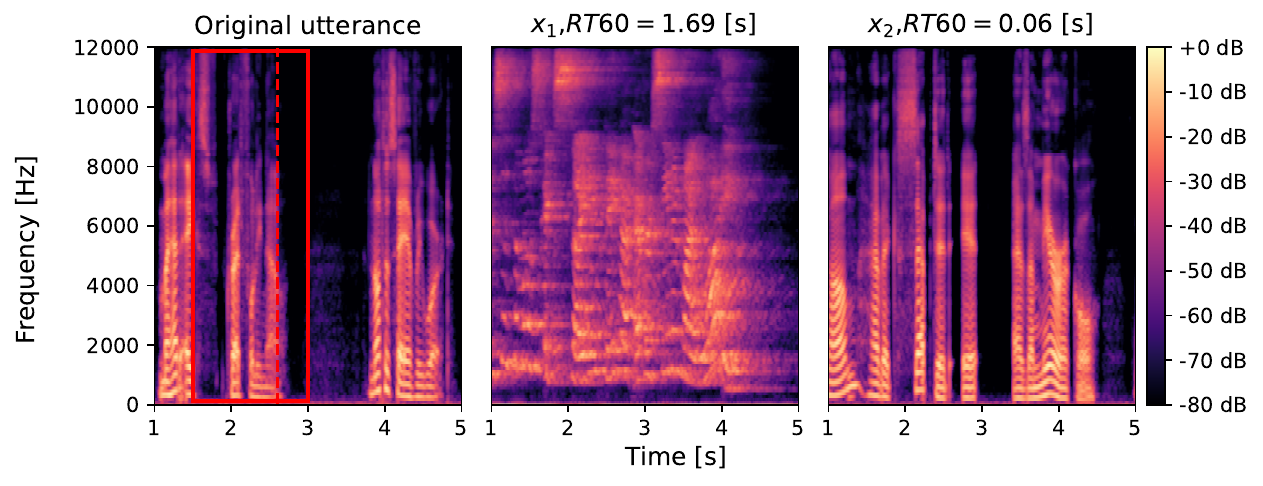}
        \caption{Original utterance, $\bm{x}_1$ and $\bm{x}_2$ used for reverb feature interpolation. $\bm{x}_1$ corresponds to reverberant speech ($\text{RT60}=1.69$ seconds) and $\bm{x}_2$ corresponds to the non-reverberant speech ($\text{RT60}=0.06$ seconds). Red box corresponds to the zoomed-in section in \cref{fig:interpolation}}
        \label{fig:sources}
    \end{subfigure}
    \begin{subfigure}[t]{\linewidth}
        \centering
        \includegraphics[width=0.87\linewidth]{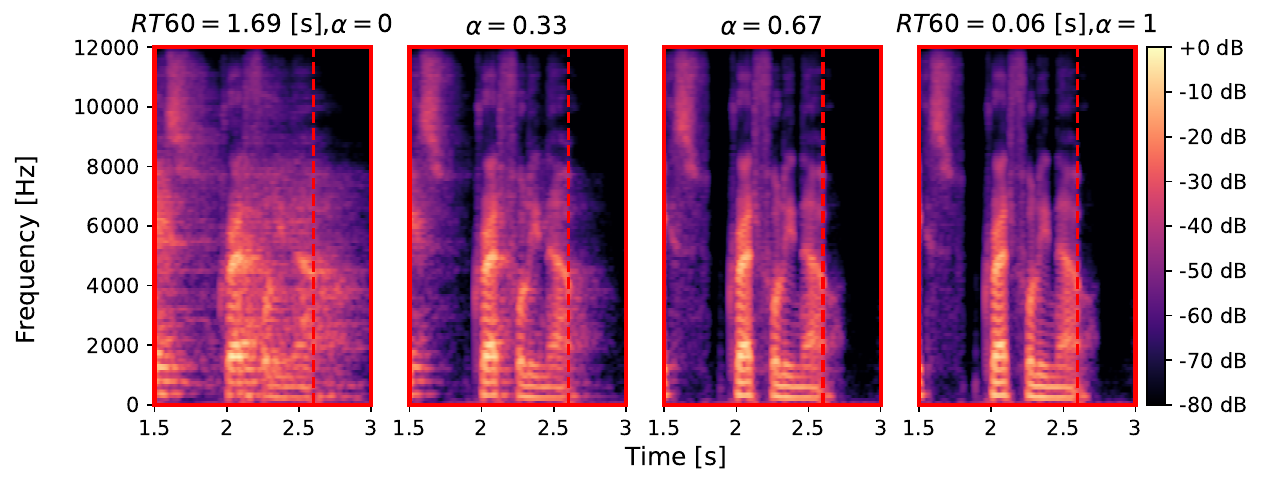}
        \caption{Zoomed-in visualization of reverb feature interpolation. Red dashed line shows the part where the reverberation change is clearly visible.}
        \label{fig:interpolation}
    \end{subfigure}
    \caption{Reverb feature interpolation inputs and corresponding output with different $\alpha$.}
\end{figure}
\begin{figure}
    \centering
    \includegraphics[width=0.87\linewidth]{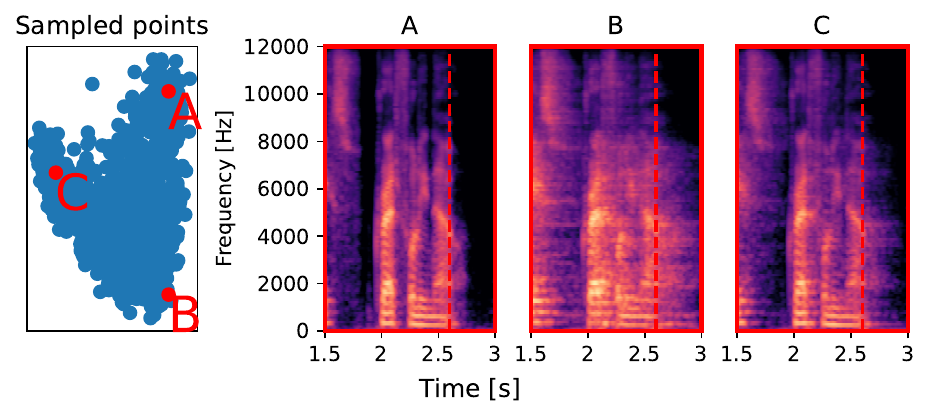}
    \caption{Reverb feature sampling from PCA-based latent space. The original utterance is the same as the \cref{fig:sources}}
    \label{fig:sampled}
\vspace{-0.4\baselineskip}
\end{figure}

Given the observed continuity of the reverb feature concerning reverberation parameters such as RT60 and DRR (\cref{fig:reverb_feature}), linear interpolation presents a viable method for generating novel reverberant effects. 
Let $\bm{c}_1$ and $\bm{c}_2$ be a reverb features extracted from uttrances $\bm{x}_1$ and $\bm{x}_2$. Interpolated reverb feature, $\bm{c}(\alpha)$, can be computed from $\bm{c}_1$ and $\bm{c}_2$, using a weight $\alpha$ as 
$
    \bm{c}(\alpha) = (1-\alpha) \bm{c}_1 + \alpha \bm{c}_2,\text{for } 0 \leq \alpha \leq 1
$.
\cref{fig:interpolation} illustrates the result of applying interpolated reverb features to speech synthesis, visualized via mel spectrograms. For this demonstration, $\bm{c}_1$ corresponds to a highly reverberant condition (RT60 = 1.69 s) and $\bm{c}_2$ to a near-anechoic condition (RT60 = 0.06 s) as shown in \cref{fig:sources}. The result confirms that varying $\alpha$ from 0 to 1 yields continuous modification of reverberance, while maintaining its harmonic structure integrity. This shows that it is possible to control the perceived reverberation through straightforward linear manipulation of the proposed feature. Corresponding audio samples can be found online$^\text{\ref{fot:demo}}$.

\subsection{Reverb-feature sampling}

The reverberation interpolation method in ReverbMiipher assumes the user provides reference speech with reverberation characteristics similar to the desired output, which limits its applicability. Therefore, we also explored an alternative approach where the reverberation feature is sampled directly from its latent space. Specifically, we sampled features from the two-dimensional plane derived from the aforementioned PCA analysis.
\cref{fig:sampled} presents the results of this sampling strategy. The results demonstrate that the original speech structure is preserved while the level of reverberance changes for each sampled feature. This suggests that the straightforward strategy of sampling from the human understandable 2D PCA plane is viable for controlling reverberation.
Audio samples can be found online\textsuperscript{\ref{fot:demo}}.

\section{Conclusion}
In this paper, we introduced ReverbMiipher: an \ac{SR} model designed to preserve and manipulate reverberation while removing unwanted distortions such as additive noise or codec artifacts.
ReverbMiipher extends the Miipher-2 SR model with a reverb encoder that extracts reverberation-related features from the noisy speech input.
Experimental results show that ReverbMiipher preserves reverberation better than to the baseline approach, which uses ground truth RIR reverberation characteristics for preservation.
We also demonstrated that the reverberation manipulation is possible with both reverberation interpolation and reverb feature sampling. 
Future work includes preservation or manipulation of other acoustically-encoded information, such as environmental sounds, for SR.

\clearpage
\bibliographystyle{IEEEtran}
\bibliography{refs25}
\end{document}